\documentclass[a4paper,12pt]{article}
\usepackage{amsmath}
\usepackage{amsfonts}
\usepackage{amsthm}
\usepackage{graphicx}

\begin{document}

\begin{center} {\Large {\bf Real Islamic Logic}}
   \\ \baselineskip 13pt{\ } \vskip 0.2in
\large{Jan Aldert Bergstra}\\~\\
University of Amsterdam\footnote{Informatics Institute, Section Theory of Computer Science, 
Faculty of Science, University of Amsterdam, The Netherlands. Email: j.a.bergstra@uva.nl.

I acknowledge several discussions with M.P. Lindo (University of Amsterdam), who in particular drew my attention
to \cite{Maurer2005} which provides a picture of Islamic Finance from an anthropological perspective.}
\\(\today)

\end{center}

\bigskip

\begin{abstract}
\noindent
Four options for assigning a meaning to Islamic Logic are surveyed including a new proposal for an option 
named {\em Real Islamic Logic}  (RIL). That approach to Islamic Logic should serve modern Islamic objectives in a way 
comparable to the functionality of Islamic Finance. 

The prospective role of RIL is analyzed from several perspectives: (i) parallel distributed systems design, (ii) reception by a community structured
audience, (iii) informal logic and applied non-classical logics, and (iv) (in)tractability and artificial intelligence.
\end{abstract}

\section{Introduction}
Islamic Logic is often used as a label for Arabic logic which flourished between 900  and 1500. Indeed a rich literature
on logic and philosophy has been produced which merits descriptions like Arabic Logic, Arabic School of Logic, and perhaps less 
plausible Islamic Logic.
Many scholarly works have been written about that period and aspect of Arabic culture as well including the reception of Aristotelian logic.

Using Islamic Logic primarily as a pointer to a part of the history of logic is in marked contrast with the common usage of the 
phrase  ``Islamic Finance'', nowadays mainly perceived as an indication for a modern and thriving phenomenon which must not be understood through its centuries old historic roots but instead should preferably be understood as a form of Jihad.\footnote{I will assume that a manifestation of Jihad consists exclusively of non-violent actions by default. Manifestations of violent Jihad will play no role in the paper. 
A proper understanding of Islamic Finance a as a part of political Islam cannot be obtained without some awareness of the history of 
Islam from say 1920 onwards.} 

Consequential to the reception of Aristotelian logic, the logic of syllogisms has been criticized by Islamic scholars. The ample existence of this
criticism may lead one to believe that Islam has been and perhaps still is rather pessimistic about logic.\footnote{The judgement that early 
and later Islamic scholars considered logic forbidden ({\em haram}) has been put forward to great effect by Goldhizer in \cite{Goldhizer1915}. Recent work by Khaled El-Rouayheb \cite{ElRouayheb2004} and Spevack \cite{Spevack2010} indicates that at no time the opponents of 
logic had a significant majority.} That far reaching conclusion, however, 
becomes less defensible if one imagines that these critical scholars simply saw no plausible way for Greek logicians to arrive at the 
universally quantified assertions from which they went onwards with reasoning by means of instantiation as incorporated in various 
syllogisms (see Ruth Mas \cite{Mas1998}). Universally quantified assertions cannot be ``sensed''  and for that reason they can be 
constructions of the individual mind only, thus lacking the required ``objectivity'' so it was argued. That a modern scientific process 
may generate universally quantified assertions which are considered objectively valid was unknown in the days when syllogistic logic came under attack of Arabic scholars. Their criticism need not be construed as an outcome of religious contemplation. Fortunately (for Islamic scholars)  the revealed sources, which were unaccessible to ancient Greek scholars,  codified after having been  made available through prophecy, could produce fully reliable universally quantified assertions. This observation, perhaps more adequately understood as a design decision about the basis of logical reasoning, then provided a clever way to solve a pressing problem in the philosophy of science, but nowadays it represents an outdated view. A modern perspective on science, with its method for generating universally quantified statements acceptable for entire communities rather than for single individuals only, implies that formalized logic can be accepted more easily, and implies that  the mechanics of reasoning need not be made exclusively dependent on the existence of revealed sources. Summing up these arguments I see no reason to believe that Islam is nowadays intrinsically hostile towards methodical reasoning on the basis of large collections of assumptions,  neither do I believe that Islam is committed to the viewpoint that only assertions obtained directly or indirectly from 
revealed sources can be both universal and valid. In addition there is no compelling reason why a systematic investigation of reasoning processes (that is logic as a branch of mathematics, science and philosophy) should be considered incompatible with Islam.
Nevertheless the  contemporary tolerance of Islam towards logic seems not to have given rise to any specific developments that
may foster the usability of logic for Islam. 

Most developments of applied logic remain within a limited set of domains: computing (very significant), mathematics (limited), physics (very limited), economics (marginal but growing alongside the explosion of electronic trade), law (still marginal but steadily growing), linguistics (large but perhaps on the way back), psychology (marginal and perhaps even declining), and philosophy (marginal and stable). 

Now there is no technical reason why applications of logic to religion and in particular to some specific religion cannot be developed. Of course a logician may be unwilling to work towards progress in that direction, but however distant religion may be considered from science that distance by itself
cannot rule out a potential for applications of logic in that direction.\footnote{Applications of logic to a system of law can be imagined irrespective of the moral quality attributed to that system, though one's willingness
to perform research towards that end may very well depend on the perceived moral quality of the particular system of law at hand.} This paper analyses a perspective for an application of logic within Islam from a position outside Islam.\footnote{I see no compelling argument against this objective stemming from implicit or even explicit intrinsic ethical problems which might be attributed to Islam from an external position. Proving decisively that such objections do not exist is quite another matter of course.}

\subsection{Logic applied to a specific religion}
If one accepts that logic may be applied to or within a religion in spite of the fact that logic is a constituent of science and philosophy whereas 
religion is not, there are still many ways in which this kind of application can be conceived. I will refrain from an attempt to survey these options and focus on one particular view only which likely to be specific for Islam.

Working towards a definite and applicable form of Islamic Logic may be conceived of as an instance of Islamization of modernity, with conventional formal logic considered modern. Islamization of modernity also underlies Islamic Finance. Through Islamization of the logic of legal argument, and successive application of that logic for structuring legal reasoning, legal arguments may become more transparent and more convincing, thereby promoting underlying values of Islam in a novel fashion. In this way Islamic Logic may be considered a potential manifestation of Jihad.\footnote{Perhaps Jihad may be translated as ``the progressive movement from within Islam'', where ``progressive'' still may have many mutually highly incompatible meanings.}

\subsection{Usage of language}
In this text I will avoid Arabic terms as much as possible based on the hypothesis that the fundamental content of Islam is international and language independent. Unavoidably translations are used, for instance instead of Shari'ah I will use Islamic Legal Process, and instead of Allah I will write God.\footnote{This convention is for instance followed by Makdisi in \cite{Makdisi1985} but is is unusual in writings on Islamic Finance. Writing Allah instead of God in an English text may be misunderstood as a suggestion that these are different entities after all, thereby potentially contradicting the principled monotheism of Islam. I will not make use of a translation of the term {\em Islam.} Instead an attempt is made in this paper to provide a working definition of Islam. The translation ``submission (to God's will)'' as given by Wikipedia can be used for the purposes of this paper, however.}

\subsection{The problematic status of speculative thought}
Writing this paper has proven difficult for various reasons. First of all there is no way that I can take into account or even become aware of the enormous literature which has been accumulated regarding the topics that will be covered. Now specialization is a major tool for fighting the enemy of amateurism in research, and reliable claims of novelty are usually based on equally reliable accounts of relevant prior art. But I feel the need to arrive at a comprehensive picture and the objective to achieve some form of completeness when writing this paper has taken priority over scholarly precision concerning the search for backwards pointers to prior art. 

Moreover there is a speculative aspect about any development aiming at relating a classical part of science and philosophy (in this case logic) with any form of religion.  If one intends to argue that such connections may provide an advantage the story may become even more speculative. Most work on Islamic Finance and Islamic Logic that I have seen avoids speculation by making use of a descriptive style: historical work describes what people did in the past and how we may find out about that, anthropological work looks at how small communities deal with some body of methods and concepts, sociological work considers the development of communities at large and economic work takes measurable streams of goods, products and services into account. I try to write about a modernization of Islamic logic with the additional constraint that, although being an outsider, I need to understand 
and trust the full story including a rationale of Islam and the role that logic may play in that context. 

\subsection{About the external position}
Being an outsider to Islam, and writing from that position, must not primarily be valued as providing a tool for arriving at an objective perspective and for avoiding speculative thoughts.\footnote{An informative reference concerning the issue of taking an external position, while crossing borders every now and then can be found in a book review by Victor Kal \cite{Kal2002}. Another way to make use of an external position is found in \cite{DarPresley1999}, where extensive suggestions are made how theorists of Islamic Finance can make better use of existing ``western'' economic theory. The authors phrase these suggestions as an advice to ``Islamic Scholars'' and do not specify to what extent this advice might be taken up in their own work for instance.} Rather taking the external position reflects a fact of the matter which I must in addition take into account without knowing in advance to what extent it simplifies or complicates the task at hand. Working from inside Islam and working from outside Islam provides an author with different interfaces of operational options. An author cannot freely choose and change his position in this respect, as if taking an inside Islamic perspective (from an outsider's initial position) were conceivable as merely a tailor made and methodologically legitimate anthropological style of work, which can be dispensed with after the planned research activity has been completed, thereby returning to the initial outsider position.

An undeniable obligation for an outsider is to be sufficiently critical. Have I overlooked intrinsic problems of Islam (seen from an outsider's perspective)  that should have been mentioned in this paper? Am I participating in not seeing what needs to be seen? Of course I don't think so myself. I will return to this difficult issue in the concluding section of the paper. The computer scientist C.A.R. Hoare wrote: 
\begin{quote}
inside every large program is a small program struggling to get out.
\end{quote}
This remarkable observation is meaningful (and useful) even if the large program contains many design and programming errors. Transposing Hoare's observation from the science of computer programming to the study of  religions I obtain: 
\begin{quote}
inside a complex religion is a clear core struggling to get out and to become visible and comprehensible from an outsider's position.
\end{quote}
This holds true to a considerable extent  in spite of mistakes enacted in the name of that religion. Thus the outsider's position to a religion may be compared with the informed programmer's view on a large program written by an unknown software engineering team: find the core which is struggling to get out, and don't be distracted from that task by the discovery of programming errors because the existence of programming errors is a statistical fact which cannot undermine the validity of Hoare's observation.

\subsection{Novelty remains to be seen}
No claim of novelty of any assertion put forward in this paper can be reliably based on the author's grasp of the body of published and grey literature relevant for that particular assertion. Even in very confined areas of mathematical logic and the theory of computing it has proven strikingly difficult to find out what has been done by other authors before.\footnote{Our recent work on proposition algebra and short-circuit logic (see \cite{BergstraPonse2010b} and further work cited in that paper) has been a remarkable experience concerning the difficulty of surveying prior art. Working with 3 to 4 persons quite systematically it still took us almost two years to spot an essential (but nowhere cited) 45 years old paper by Tony Hoare. And of course that finding alone comes nowhere near completing the prior art discovery job that needs to be done just for that particular and quite specialized topic.}

To the best of my knowledge the proposed description of Real Islamic Logic is new in objective, in form, and in content. The stratified description of Islam seems to be new. The focus on autonomous  parallel legal processing as a key strength worth of preservation and enhancement  of the Islamic Legal Process may be new. The excursion in the final section towards an additional (new) argument for assuming the existence of God may also consititute an original contribution.

\section{Four conceptions of Islamic Logic}
Islamic logic can be understood in four different ways at least, the historic conception and the holistic conception 
being the only views that we found in the existing literature. These four understandings of Islamic Logic will be labeled as follows: (i) historic, (ii) holistic, (iii) Real (Islamic Logic), and (iv) Crescent-Star Logic.\footnote{Crescent-Star Logic may be considered an abbreviation of Crescent-Star Islamic Logic, which is a topic that may be technically  treated without any mention of Islam but for which the origin and motivation can only be found by taking Real Islamic Logic into account.}

\subsection{Historic topic}
The dominant understanding of Islamic Logic is that this refers to a period in the development of logic. Pinpointing that period is rather arbitrary, but 900-1600 is a reasonable guess. It might alternatively, and according to some authors more appropriately, be called Arabic Logic. 
The term may be compared to: ``Roman Architecture'', ``French Literature'',  ``Greek Philosophy'', or ``Welsh Poetry''. The historic understanding of Islamic Logic is so widespread that it may be considered the standard meaning of the phrase. Here are some pointers to the rich literature on the subject:
\cite{AlwishahSanson2009}, \cite{Sabra1980}, and \cite{Street2000}. Islamic Logic is less formal than present day manifestations of logic. We find that what Islamic Logic as accumulated some 1000 years ago is comparable to what today is called informal logic rather than formal logic or to philosophical logic.

It should be noticed that Islamic Finance seems to have no historic connotation, whereas for instance Islamic Architecture does. Islamic Architecture may be decomposed into historic and modern. Modern Islamic Logic currently seems to be a vacuous concept at present.

\subsection{Holistic conception}
Islamic logic is sometimes understood as the way muslims are supposed to think. Such applications of the term are always implicit, never based
on a scholarly analysis and may be considered a misuse of the term logic. This form of (mis)use is rather widespread. For instance ``Capitalist Logic'' may be called into action as a rationale for the dictate that minimal wages must be paid and so on. The phrase ``Economic Logic''  is often used  with a similar holistic interpretation. Its use is almost never convincing and nearly always one might ask: which economic logic is meant in particular. There is no useful connection between the holistic use of logic and logic as a theme in philosophy, mathematics or science.

\subsection{A manifestation of Jihad: Real Islamic Logic (RIL)}
Novel as far as I know, but nevertheless potentially attractive, is to consider Islamic Logic as a  longstanding ``project''  for the 
advancement of Islam: 
Real Islamic Logic. Real Islamic Logic embodies the Islamization of Logic (as a part of modernity) just as Islamic Finance embodies the Islamization of finance. The adjective real is added in order to avoid confusion with the three other conceptions of Islamic Logic. It is conceivable that after due time the ``real''  can be left out and Islamic Logic has undergone a change of its default meaning.

Below a proposal will be made concerning the table of contents and the repertoire of objectives of Real Islamic Logic. The proposal may be read as a proposed political view on  {\em how to advance Islam by way of the Islamization of Logic}. In this conception of Islamic Logic  the similarity with Islamic Finance is taken to a rather extreme conclusion and all, often implicit, political objectives that underly Islamic Finance are considered 
candidates for being taken into account when developing a specification in detail for the design and development of Real Islamic Logic. If it ever gets off the ground the project of developing a Real Islamic Logic may take centuries rather than decades. 

(Real) Islamic Logic can be conceptualized by a non-Islamic author, but it cannot be performed to its full potential without being part of the Umma. For the (non-Islamic) author of this paper this limitation implies is that he can carry out conceptual work towards the establishment of the Islamization of logic interpretation of Islamic Logic (that is RIL), but he must stop at its doorstep once it has been designed.\footnote{Reasons for a reluctant approach in bringing RIL into practice are twofold. (i) It is implausible that a non-muslim performs actions that may be classified as Jihad, and, (ii) such actions can be strongly criticized by non-muslims because of perceived ethical problems. I will expand on this matter in section \ref{MoralDs}.}

It is not so easy to find a name for an activist and contemporary interpretation of Islamic Logic. I am suggesting to use Real Islamic Logic, thus recognizing and leaving untouched, the currently predominant historic interpretation of the phrase ``Islamic Logic''.\footnote{Retrospectively that leads to Real Islamic Finance as a name for Islamic Finance useful for those who would prefer Islamic Finance to have a historic connotation just like Islamic Logic. However, there are reasons not to identify Real Islamic Finance with the current meaning of Islamic Finance and to keep room for a forthcoming specification of Real Islamic Finance which expressively does away with some of the historic imports that might have been needed for getting Islamic Finance moving but which may in the mean time have become a drawback. More specifically Real Islamic Finance might be conceived as being more flexible towards interests being paid and received than Islamic Finance claims it must be on theological grounds.}

Stopping at the doorstep of Real Islamic Logic may be considered frustrating for a non-muslim intrigued by the subject. This can be solved by recasting the subject in non-Islamic terms. Facilitating that step motivates the fourth and final understanding of Islamic logic.

\subsection{Cresent-star Logic}
Islamic Finance can serve as a role model for the development of Real Islamic Logic. In both cases the projection to a non-religious version of the topic can be performed in a similar fashion.
In \cite{Bergstra2011b} I have suggested Crescent-star Finance as a reference to Islamic Finance explicitly stripped from its Islamization of finance perspective. Crescent-star Logic casts Real Islamic Logic  as a neutral topic in logic stripped from  the Islamization of modernity ideology and stripped from the Islamic tradition.\footnote{Real Islamic Logic is supposed to have been detached from the Arabic tradition already so that step need not be performed when making the move towards Crescent-star Logic.} Crescent-star Logic does away with the conventional historic perspective of the scholarly interpretation and does away with the Jihadic perspective of the Islamization of modernity.\footnote{It is not meant that Islamization of modernity is the only or even a major position of Jihad towards modernity. It merely constitutes an option that can be considered relevant in the case of logic.} Crescent-star logic then represents the pure logic of Real Islamic Logic as formulated above. Due to its neutral form Crescent-star Logic should be
accessible and legitimate for logicians who oppose the very concept of Islamization of Logic, assuming that the mere perspective of pure logic in 
an ideologically biased praxis presents no unsurmountable obstacle.

Progress in Crescent-star Logic may help to improve the usage of logic for the application in Islamic courts. Crescent-star Logic 
uses and integrates several logical themes which must come together forming a complex of reasoning methods of uncommon complexity:
\begin{itemize}
\item paraconsistent logic (local inconsistencies cannot be excluded, not even in the case of a single court),
\item deontic logic (most judgements/assertions are about permissions and prohibitions),
\item belief revision (courts may withdraw earlier assertions representing court beliefs that have become superseded),
\item modal logic (in some cases different courts may best be seen as logically existing in different worlds).
\end{itemize}

\section{A community oriented description of Islam}\label{defOfIslam}
Who belongs to Islam, and what themes and activities can be called Islamic? These questions are supposed to have some kind of an answer as soon as one starts writing about Islamic Finance or about Islamic Logic. 
The simplest option is to assume that Islam is a well-known notion provided with useful definitions elsewhere so that the matter need not be discussed in further detail. That assumption seems to underly much writing in these areas but I doubt its validity. Below I will provide a description of the extension of Islam which yields a reasonably explicit criterion allowing to assess to what extent an approach to finance may be called Islamic and to what extent a contemporary approach to logic deserves being labeled (Real) Islamic.\footnote{This description has been derived from the sociology of computer science which nowadays is entirely community based so it seems.}

It is assumed that at any moment of time Islam consists of a collection $I^t$ of persons with 720  $\leq t$. This set $I^t$ needs to be identified for various $t$. A sequence of 8 sets of persons $I_0^t, I_1^t, I_2^t, I_3^t, I_4^t, I_5^t, I_6^t$  and $I_7^t$ will be defined, each less inclusive. These sets change in time, for instance because the deceased must be deleted from each set in which they have been included at the time of their death. New persons can enter for the first time at all stages except stage 6 and 7. Only death removes persons from the set $I_0$. Members can move up and down through the various levels this hierarchy so to speak.

By means of this sequence a reasonably precise definition of who belongs to Islam at time $t$ can be provided. This nested series of sets can also be used to determine when an activity can be called Islamic. Thus at any instant of time $t$ the sets have a specific extension $I_n^t$ with $0 \leq n \leq 7$ each consisting of persons living at time $t$, with $I_n^t \subseteq I_{n+1}^t$ for $0 \leq n < 7$.

At any time Islam will mean one of the sets $I_n^t$, however with $I_5^t$ constituting the default extension of Islam. If another extension is meant that must be mentioned explicitly. 

Why is this relevant? For instance, even after having read extensively about Islamic Finance one may still wonder:
what justifies the label Islamic for these financial activities?\footnote{And similarly one may wonder with what justification some activities are ascribed to ``Islamic extremism'' (assuming that it is known ``who did it, and why'').} Can someone, at least in principle, define his or her
 own system of Islamic Finance (or of Islamic Logic) as a theoretical project in a Northern European University  carried out by non-muslims, or is such a state of affairs impossible by definition? 

None of these questions can be given a reliable answer if no extension of Islam and of points of view ``held by Islam'' is known, and the objective of this section is to make some progress on that matter. Unfortunately but not uncharacteristically we will make use of sources that are not full in agreement with one-another. Except for the design of the sets $I_{0-7}^t$, and except for issues concerning Islamic Finance, the knowledge of Islam used in this section has been taken mainly from the following sources Armstrong \cite{Armstrong2001}, Donner \cite{Donner2010}, and Esposito \cite{Esposito2010}.

Donner's book, proposes an intriguing and attractive picture of the prophet Muhammad and his contemporaries, organized as a community of believers, until the creation of Islam some 80 years after the death of the prophet. He departs from the conventional picture as sketched by Armstrong. Esposito forcefully argues against North American prejudice. He bases his arguments on several international survey studies, conducted by Gallup, about the opinions held by members of Islam regarding a range of issues. In terms of the sequence $I_{0-7}^t$ that will be defined in more detail below, I guess that Esposito describes the result of interviews of members of the stage   $I_5$ in that listing. Supporters of the 9-11 attacks have been included in the reported polls, but there is no quantitative information provided in \cite{Esposito2010} about the coverage of the distribution of questionnaires over the different streams within Islam that Esposito intends to distinguish.

\subsection{Extension independent aspects of Islam}
As stated above our objective is to shed light on the following: who belongs to Islam (the extension of Islam), what are Islamic points of view, what qualifies an approach or method as Islamic. We begin with an ``axiom''.
\\~\\
\noindent AXIOM 1: What Islam has to say about a topic $T$ is entirely and exclusively determined by the points of view of those persons who currently are to be considered as belonging to Islam. 
\\~\\
\noindent This axiom has several implications which are rather independent of time and for that reason from the actual extension of the membership of Islam:
\begin{itemize}
\item Islam's view concerning $T$, if it exists at all, can change over time.
\item Islam's view on $T$ cannot be discovered exclusively by reading old sources.
\item During the life of the prophet Muhammad Islam did not yet exist, the Qur'an for that reason, assuming that it coincides with Muhammad's spoken words, is not an immediate source for ``Islamic viewpoints''.
\item In most cases (that is for most $T$)  some kind of ``voting'', either explicit or implicit will be required to arrive at an Islamic viewpoint about topic $T$.
\item Islam is entirely man made, even if its sources may have been be revealed. 
\item Confirmation of the revealed status at time $t$ of the original sources is part of becoming a member of the set $I_{0}^t$
\item The construction of Islam is an ongoing process with revealed sources and their continuous interpretation playing a very important role.
\item The elevation around the year 720  of Muhammad's oral tradition, after its written compilation (perhaps around the year 650), to the status of the
 primary revealed source of Islam has been an extremely successful conscious design decision that went into the construction of Islam. Further successful design decisions were to come.
\item Design decisions about Islam are exclusively taken by members of Islam. Such design decisions are just points of view about certain topics turned into assertions that must be confirmed by new members of stages 5 and 6. In terms of the hierarchy of sets such a decision can adapt the bundle of points of view adhered to by mainstream communities and for that reason incorporation of a decision may reduce the number of mainstream communities (thus moving their members back from level 6 to level 5.)
\item Who belongs to Islam, is exclusively decided by members of Islam, and this cannot be undone retrospectively, that is who belongs to Islam at time $t$ will always belong to Islam at time $t^{\prime} > t$ though perhaps at a different level of the stratification.
\end{itemize}

\subsection{A stratified membership description}
In order to gain insight in the extension of Islam at an arbitrary moment in time an axiom is used that allows for a layered decomposition of its membership.
\\~\\
\noindent AXIOM 2. Islam provides unity in diversity. For membership of Islam distinguish 7 levels can be distinguished, 
which are incrementally more demanding.
\\~\\
\noindent ABBREVIATION:  $V$ is used as an abbreviation of the following vow: ``God is the only god and Muhammad is his most prominent and most recent prophet''.

\begin{description}
\item{\em Retrospective subjective members.} $I_0^t$ consists of those (living) people who have at some time $r \leq t$ expressed $V$ (with or without the presence of witnesses).

According to some $I_0^t$ coincides exactly with members of Islam at time $t$. What can be held against this viewpoint is that persons may be insufficiently aware of the consequences when making the vow $V.$

Members of $I_0^t$ may be members of other religions as well (usually only after revoking vow $V$).\footnote{There is some confusion that Islam would not tolerate persons revoking $V$. Revocation is often discouraged but it cannot be prevented. Different communities within Islam may have different viewpoints concerning the legality of revocation and concerning its consequences.}

\item{\em Subjective members.} $I_1^t$ consists of the members of $I_0^t$ who have not revoked their vow $V$ since asserting it.

$I_1^t$ is disjoint with Judaism, all Christian religions, and with most religions from India, Japan, China.
$I_1^t$ is a well-known separator in topological terms and it can be used as a definition of Islam if atheistic (who do not accept the concept of god) and zerotheistic (who acknowledge the concept of god but are in addition of the opinion that currently no god exists) persons are 
not taken into account.

\item{\em Conscious subjective members.}  $I_2^t$ consists of the members of $I_1^t$ who are at time $t$ willing to renew the vow $V.$

\item{\em Active members.} $I_3^t$ consists of the members of $I_2^t$ who (at time $t$) perform conventional religious tasks (regular prayer,  regular gifts to the poor, making a journey to Mecca),

\item{\em Community members.} $I_4^t$ consists of members of $I_3^t$ who perform their conventional religious tasks in the context of and in accordance with a community of persons all members of $I_3^t$, Thus $I_3^t$ is the union of a collection of communities each made up from members of $I_3^t$.

No attempt is made to decompose communities into subcommunities. Doing so may be important for various purposes. It leads to a partially ordered refinement of the proposed stratification.

\item{\em Traditional community members.} $I_5^t$ consists of the members of those communities that constitute $I_4^t$ of which the 
members are (collectively) aware of:
\begin{itemize}
\item a package of viewpoints collectively considered a consequence of Islam (though in fact often only in their specific community),
\item a line of descent in terms of communities from the initial phase of Islam,
\item a line of descent in terms of packages of viewpoints (the community lineage consists of a sequence of communities and intervals of their existence; it must be equipped with a package of viewpoints held during each of these phases). This theological lineage must have significant explanatory value for the community's current positions.
\end{itemize}

\item{\em Mainstream traditional community members.} $I_6^t$ comprises (the union of) a collection of communities who have decided that they are among the mainstream communities accounted for in stage 5, while other communities have been left out. The different communities at this level may share:
\begin{itemize}
\item some beliefs (neutral assertions),
\item some elements of orthopraxy,
\item some objectives concerning the preferrered development towards a next stage of the community's existence.
\end{itemize}
For a reliable demarcation of $I_6^t$ as a subset of $I_5^t$ it is necessary that the entire genealogy of branching communities composing $I_5^t$ is assessed with a degree of centrality within Islam. This is a matter of sociology and group structure and group interconnection analysis rather than a matter of deciding about the centrality of viewpoints. Anyhow, $I_6^t$ singles out the mainstream communities from the marginal, excentric and extremist communities.\footnote{The step from stage 5 to stage 6 is potentially subjective and risks being unacceptably sensitive to one's opinions independently of being outside or inside Islam. Nevertheless social network analysis and the recognition of scale free networks carry some promise of rendering the selection of the mainstream communities from stage 5 objective.}

\item{\em Mainstream based forward movers.} $I_7^t$ consists of selected members of various communities existing at level $I_6^t$, who actively perform Jihad and in various ways are quite visible at least within the membership of $I_5^t$ and perhaps also outside $I_5^t$ and even outside $I_0^t$. Here Jihad may take various forms, for instance:
\begin{itemize}
\item spreading the word,
\item living a life with visible piety, which may be convincing for others.
\item developing innovative activities that allow groups of persons to perform their life compatible with a package of viewpoints held by their community as mentioned in the specification of $I_5^t$.
\end{itemize}
\end{description}

\subsection{Qualification of views and activities}
Having dealt with the classification of individuals and groups the classification of activities and viewpoints can be put on a reasonably firm footing.\footnote{The number of stage that is taken into account in the stratification of Islam determines a level of resolution that can be achieved for the classification of activities and viewpoints.}
\\~\\
\noindent DEFINITION 1. An assertion belongs to the points of view of Islam if some community as mentioned in the definition of $I_5^t$ adheres to the assertion.\footnote{Anthropological work by Bill Maurer \cite{Maurer2005} indicates that muslim communities are willing to grant the label Islamic to activities which are in conformance with the requirements that these communities deem necessary even if that conformance is not intentional. So in some cases our definition will not provide a justification of what takes place. Perhaps definition 1 can be extended to include those activities that are considered Islamic by a stage 5 (or higher) community.}
\\~\\
\noindent DEFINITION 2. An assertion belongs to the points of view of mainstream Islam if some community as mentioned in the definition 
of $I_6^t$ adheres to the assertion.
\\~\\
\noindent DEFINITION 3. An assertion belongs to the shared points of view of Islam if a large majority (over 75\%) of communities as mentioned in the definition of $I_5^t$ adheres to the assertion.
\\~\\
\noindent DEFINITION 4. An assertion belongs to the shared points of view of mainstream Islam if a large majority (over 75\%) of communities as mentioned in the definition of $I_6^t$ adheres to the assertion.
\\~\\
\noindent DEFINITION 5. A theory, methodology or system can be called Islamic if it is endorsed by at least one community included in $I_5^t$. It is a mainstream Islamic theory, method or system, if it is endorsed by one of the communities constituting $I_6^t$. It is shared (or shared mainstream) if it is endorsed by a significant (over 75\%) of communities of $I_5^t$, respectively of $I_6^t$.
\\~\\
Here are some examples of the use of these definitions.
\begin{itemize}
\item That a woman should not be driving a car is a point of view of mainstream Islam.
\item That a woman is allowed to drive a car is a point of view of mainstream Islam.
\item That drinking alcohol must be avoided is a shared point of view of Islam.
\item That non-Islamic (or rather outside $I_2$) western civilians can be aggressively attacked is a non-shared, 
non-mainstream point of view of Islam.
\item That the poor should be supported is a shared point of view of Islam.
\item That interests on loans should neither be paid nor collected is a non-shared point of view of mainstream Islam. The support for this point of view is growing, however, and it may well become a shared point of view of mainstream Islam in the next 100 years.
\end{itemize}

Some further  consequences of the stratified definition of Islam can be mentioned:
\begin{itemize}
\item from outside $I_2^t$, on cannot design a system of Islamic finance, by definition, unless appropriate endorsment is obtained. (This is the impact of definition 5.)
\item The set of points of view of Islam is inconsistent, mainly because viewpoints that stem from more than one community are taken into account.
\item In fact even the family of shared points of view of mainstream Islam is not protected against logical inconsistency. But that is a matter related to the fact that majority voting processes can lead outside logical validity, which is generally a fact of social choice theory unspecific for Islam.
\item Given this layered architecture of Islam it is not even difficult to design (by way of a thought experiment) 
a package of LinkedIn groups which allow to capture all of Islam as well as the dynamics of its constituting communities as dedicated (an perhaps preferably closed) LinkedIn groups. That requires 1.500.000.000 persons to be covered by LinkedIn, which is rapidly becoming technologically feasible. Leaving aside legal and political objections before long large movements like Islam can be entirely covered by social media, at least in principle.
\item Islam has held inconsistent views since its earliest days. Both Christianity and Islam may be in part understood in terms of their coming to grips with an event of political assassination of a major figure (Jesus, and Uthman respectively). A major distinction between these religions  arises, however, from that fact that Islam has kept both sides of this deep moral dilemma\footnote{This dilemma arose before the establishment of Islam thus creating the split between the Shia and Sunni movements.} within its ranks (an even within the mainstream communities composing stage $I_6^t$), whereas all of Christianity has taken side for the victim of the assassination so to speak. 

\item The logic used by Islam, if any, is a paraconsistent propositional logic. Many forms of paraconsistent logic have been developed since 1900 and ill require an extensive study to find out which version fits best.\footnote{From the extensive literature on paraconsistent logics I mention
Avron \cite{Avron1991}, Priest \cite{Priest1979}, and the relevance logics of Anderson and Belnap \cite{AndersonBelnap1975} which may be considered paraconsistent logics as well.}The Roman Catholic Church on the other hand has made an attempt to live up to a consistent logic. For the union of Christian churches and their collective points of view finding a paraconsistent logic is also the best one can hope for, while for particular churches the search for full consistency has been important. That has led to fragmentation and diversity in Christianity to which Islam seems to be less prone due to its principled compatibility with local inconsistencies, that is by its ability to settle for mere pragmatic paraconsistency. Paraconsistent logics are not mainstream in the west. This is remarkable, because removing local inconsistencies is as difficult as anywhere else. I believe that ``western'' scientific paradigms tend to develop into completions of consistent subsets of originally paraconsistent theories, this process leading to a fragmentation not unlike the religious fragmentation just mentioned. 

Now we all know that if a mother asserts that her child $C$ needs to sleep at 8 PM while its father insists that it may stay awake until 9 PM both may successfully and consistently agree that it must wake up next morning at 7.30 AM, expecting that the latter goal will be aimed at unconfused by the parental disagreement about the preferred timing of $C$'s going to bed. Making sense of this situation requires a paraconsistent logic of parental behavior, however.

\item Finally the classification mechanism can be applied to the main theme of this paper. 
\begin{itemize} 
\item Doing applied research on a particular system of  Islamic Finance constitutes an Islamic activity and it requires being a member of stage 5 or beyond at least. 
\item Making proposals for systems of Islamic Finance is not necessarily an Islamic task (that is it can be done by  individuals who are not stage-0 members.)
\item Labeling a proposed financial system as a system of Islamic Finance can only be done by a group of members of stage 5 together representing at least one of its communities.
\item Doing applied work within Real Islamic Logic is a task only accessible for members of stage 5 or beyond. 
\item Drafting proposals for real Islamic Logic is accessible to non-members of stage 0. (It is not always an Islamic activity.)
\end{itemize}
\end{itemize}

Besides questioning to what extent it is reasonable to label activities in finance and logic as Islamic, one may consider this matter from the other extreme position: which processes, tasks and entities can be called Islamic, and more generally can be labeled with a religious identity. I will dwell on that matter briefly below merely coming to the conclusion that matters are far from clear.

\subsection{Some reflection on the use of a religious adjective}
I will now confront the question whether or not a seemingly neutral theme like finance or logic might be provided with the adjective ``Islamic'' or with any other religious adjective. By way of example I will consider  the adjective  ``Roman catholic''  (in the remainder  of this section abbreviated to catholic) instead of the adjective Islamic. Consider the catholic priest John active in a rural area, who administers a parish $P$, who owns a horse $H$ and who regularly serves the mass in "his" church $C$. In addition he is in charge of school $S$. It seems rather absurd to label horse $H$ Roman catholic merely because it is owned by a catholic priest. An object or structure $X$ being catholic must say something significant about $X$ itself, merely a reference to an owner is definitely not sufficient. 

\subsubsection{Religious labeling of material objects, books, theories and thoughts}
So what about the school. If the horse cannot be catholic can the bricks and glass constituting the school be catholic? It seems more plausible to assume that the term ``school'' must be disambiguated, because it refers both to an organization, which might be labeled catholic, and to a physical building which is used by this organization, the building being less amenable to a religious adjective. For the church building the same remark may apply: as a building it is hardly amenable to an application of the adjective catholic, whereas the community making use of the church is plausibly labeled catholic. But where is this reductionist strategy leading to. The organization running the school and the community constituting the parish (and making use of the church) can be labeled catholic merely because the members of these groups are considered catholic. But that is obviously  insufficient, because if these organizations/communities are to be perceived as catholic, besides a constraint on their membership this also imposes the requirement of having acquired an adequate accreditation by the local catholic bishop, acting on behalf of the catholic pope. This brings us reasonably close to the definition of a catholic school (or parish).

One may then consider a book with catholic religious hymns used in church $C$. Is this book with hymns catholic, and if not is the collection of hymns itself catholic, or is it merely a neutral  tool for an activity performed by a catholic community. Similarly a book may contain the catechism of the  catholic church, which is not to say that the book itself is catholic. 

Finally one may face the question to what extent a body of ideas can be considered catholic. This leads to the specific question whether or not the catholic faith (as a collection of ideas) is itself catholic. If not then $X$ being catholic is not even a precondition for ``catholic $X$'' to make sense. If so, then that faith constitutes a body of ideas which is rightfully labeled catholic. Alternatively one might also hold that ``catholic faith'' rather than referring to something catholic specifies someone's state of mind which then might be considered catholic. That state of mind is not amenable to episcopal accreditation, however, neither is any other property a person inherits from his or her state of mind.

Summarizing these considerations I conclude that the question ``when can an $X$ be called Islamic'' is quite difficult to answer but this difficulty is not specific for Islam, but rather independent from Islam. Further for a specific theme or these $T$ the question ``can $T$ be called Islamic'' can be analyzed in sociological terms by making use of the layered stratification of its extension. Application of these matters is not at all obvious, however. The question ``is the Qur'an an Islamic text'' indicates some of the complications involved. The answer to this question may be negative if one thinks of Qur'an primarily in the time of its writing but it may be positive when it is analyzed in terms of its much later reception. So it appears that the latter question is insufficiently specific to allow for a definite answer.

\subsubsection{An instrumental view on the label Islamic}
The simplest way to appreciate Real Islamic Logic and Islamic Finance is to assume that logic and finance get colored in religious terms because of the intended application. An instrument used for Jihad may be labeled an Islamic instrument, even  if it might be used alternatively for the opposite purpose just as well.

This convention being somewhat unsatisfactory I suggest that an instrument might be labeled Islamic (mainstream Islamic) if the following three (four) criteria are met:
\begin{itemize}
 \item it is used for Jihad, using a very liberal and preferably non-violent interpretation of that term, and,
 \item it is specific (or has been designed specifically)  for that particular use, and,
 \item if the previous observations are confirmed by an uncontested group of leading figures in a stage 5 (or higher) community.
 \item (If the confirmation is provided by a stage 6 (or 7) community it is a mainstream Islamic instrument.)
\end{itemize}

Given this convention about using Islamic as a label, some further remarks can be made considering the plausibility of religious labeling in various circumstances, now rendered specifically for Islam:
\begin{itemize}
\item A recent copy of the Qur'an is a (mainstream) Islamic book (instrument for distributing  information, whereas its content has been transformed from non-Islamic to Islamic around the year 720.
\item At second inspection the decisive argument that horse $H$ above can't be labeled Roman catholic lies in the fact that $H$ is in no way specific for the catholic faith, although the way he is used may be dedicated towards strengthening that particular religion. This same argument generalizes to all animals.
\item A person is Islamic if he or she is a member of Islam. This is a matter of degree in accordance with the stratification.
\item No animal and no natural location can be Islamic. Except for persons only artifacts (including their abstract designs) can plausibly be labeled Islamic.
\item Islamic Finance is a plausible term  because the particular form of finance is supposed to satisfy the four criteria mentioned above. In fact it may 
be labeled a mainstream Islamic activity.
\item If the design of Real Islamic Finance is sufficiently specific and its intended application is sufficiently compatible with some form of Jihad that will validate the use of the phrase given the mentioned criteria.
\item the introduction of RIL in this paper does not qualify as an Islamic activity.
\end{itemize}

\section{Comparing Real Islamic Logic and Islamic Finance}
Before working out a specific proposal for Real Islamic Logic in some detail, that notion which is transpiring in an abstract or distant form already, 
will be compared to Islamic Finance which has been used since around the year 1930 onwards. This comparison is supposed to be helpful for developing an understanding of what RIL can be given a perspective on Islamic Finance.

After that comparison the companion notions Crescent-star Finance and Crescent-star Logic will be briefly compared.
\subsection{Comparing IF and RIL in some detail}
It has been argued that whether or not the label Roman catholic can be assigned to some concept is a difficult matter and that difficulty is similar for the adjective Islamic. Nevertheless some convention has been formulated and that convention underlies the understanding of the adjective Islamic in the sequel of this paper. I will now make an attempt to highlight  in detail some important merits and demerits of the phrase Islamic Finance while contrasting it with Real Islamic Logic.

\begin{enumerate}
\item As I have noticed already above Islamic Finance seems not to be used in existing literature with the historic bias which is dominantly  assumed for Islamic Logic.
\item I briefly consider Islamic Astronomy. That has a historic connotation by default. One might seek for a modern version of it  ``Modern Islamic Astronomy'' or even a Jihadic form `` Real Islamic Astronomy''  (or a movement with similar objectives). None of these exist because Modern Islamic Astronomy is simply Astronomy. 
\item There is no contemporary  Islamic Astronomy for the simple reason that conventional astronomy is entirely acceptable from an Islamic perspective. Real Islamic Logic may be considered meaningless in the same way as contemporary Islamic Astronomy is. Likewise some hold Islamic Finance to be a self-contradictory notion. However, these arguments are flawed because although Islam suggests no alternative ways for pursuing astronomy,  Islamic Finance definitely involves a rather specific set of financial conventions and the logic for Islamic Finance may be designed quite specifically as a customized toolbox for applications in an Islamic context. A similar argument can be put forward concerning RIL.
\item The logic involved in Real Islamic Logic can be pursued both in a philosophical style and in a more formalist style. At least for philosophical logic the potential to work in an Islamization oriented mode cannot be ruled out as easily as for instance for physics and mathematics.
\item Islamic Finance is a successful intellectual construction that could have originated outside Islam in theory (not what actually happened, it was a
mainstream Islamic development). Pursuing Islamic Finance is by definition a matter for muslims only.
\item A significant result of Islamic Finance is the appearance of Arabic terms and concepts in the financial world. A similar development is not to be expected from the pursuit of Real Islamic Logic.\footnote{It is remarkable that explanations of Islamic finance so often make use of Arabic terminology thereby implicitly suggesting that separating Islamic Finance from promoting the usage of Arabic is either impossible or undesirable.

Of course it may be disputed that a religion can be separated from the languages in which it has originated. Is it a modern feature of western Christianity that its core body of texts has been entirely translated into a number of different languages or is it a feature of Christianity that such a translation is an option at all? Or will it be concluded in the future that a Christian faith after all could not be successfully internationalized without becoming essentially weakened?}About this phenomenon the following can be said:
\begin{itemize}
\item The appearance of Arabic jargon in the financial world cannot simply be understood as a consequence of Islamization, in the same way as the appearance of English in many areas of activity  is not a symptom of Christianization. It merely indicates the importance of Arab speaking authors and financial workers in the pursuit of Islamic Finance.\footnote{In any case I will defend the position that Islamic Finance is a manifestation of the Islamic Progressive Political Process (an aspect of Jihad) which can and must be entirely detached from the Arabic language and its promotion. Stated more concretely: the constraint that {\em riba} must be prohibited in a written explanation of Islamic Finance must always be complemented with a binding (for the author) explanation of the meaning of {\em riba}. 

For instance who translates {\em riba} as illegally obtained financial income gets for free that ``{\em riba} is forbidden''. Getting from there to the observation that interest is forbidden though, is quite another matter. 

The relevance of this requirement on the usage of Arabic terms is immediately clear if one understands that morality of behavior is at stake. If, according to an author, a morally valid life cannot be led by persons unable to read Arabic texts that position should be made clear in advance, and definitely not be suggested as a corollary of a incomplete or even defective explanation of Islamic Finance.}
  
\item Islam has Arabic language as a major carrier of its cultural sources in very much the same way (though significantly more pronounced) as Christianity has (medieval) Latin as a source. I am assuming that conceptually Islam can and must be separated from Arabic.\footnote{If that assumption proves wrong the RIL ambition is either flawed or can only be produced by an author in command of Arabic.}
\item Not only language and religion must be separated but to some extent history and language need separation as well. For instance Shari'ah may be replaced by ``Islamic Legal Process'' and that replacement immediately removes potential misunderstandings. For instance one may think that extremely harsh punishments are characteristic for Shari'ah ignoring the fact that the local history of some well-known traditions of the Islamic Legal Process have developed in nomadic societies where imprisonment was not considered a practical option and punishment of an instantaneous form was more easily applicable. In different conditions, however, the Islamic Legal Process leads to different ways of dealing with undesirable behavior.
\item It is quite difficult to translate classical Arabic terms into English. For instance {\em gharrar} (accepting an excessive downside risk)  is forbidden in Islamic Finance but explanations of this limitation invariably involve digressions into the meaning of the term  {\em gharrar}. The concept of Islamic Finance should not be allowed the degree of freedom to proclaim that  {\em gharrar} is forbidden whatever it means so to say, thus leaving those who don't master Arabic uninformed about which behavioral limitation is imposed by means of this 
proclamation.\footnote{See \cite{ElGamal2001a} for a survey concerning the meaning of  {\em gharrar.} }
\end{itemize}
\item Currently Real Islamic Logic may be considered to be about as plausible or implausible as Islamic Economics with the difference that the phrase 
Islamic Economics is widely used. In spite of being often mentioned Islamic Economics has not really come off the ground except for its specialized financial branch.
\item Islamic Finance acquires significant visibility and profile from a single assumption namely the prohibition of interests. Opinions about the foundations of this prohibition vary from a fully religious grounding (promising no economic advantages compliant with Islamic  social objectives) to a fully economic grounding (expecting that this prohibition will contribute significantly to the reduction of phenomena of individual 
hardship\footnote{I refer to \cite{Lewison1999} for a clear and contemporary explanation of the problems to which a financial system based on interest payment may lead.} and of structural crisis). At a closer inspection Islamic Finance is based on a combination of restrictions amongst which interest prohibition is only the most well-known ingredient. These other restrictions are:
\begin{itemize}
\item Avoidance of excessive downside risks.
\item Non-reliance on excessive upside chances (gamble).
\item Sold items must exist at the time of transaction (but payment may be deferred).
\item All parties involved in a sales transaction must have comparable and complete information about what is being sold.
\item Parties involved in a financial transaction may not be forced into participation.
\end{itemize}
\item Real Islamic Logic has no counterpart to the dogma of  interest prohibition. That is no single element carries significantly 
more visibility than other elements. If any counterpart to the above concise specification of Islamic Finance must be found it 
consists of a cluster of elements for which the following listing may be a candidate:
\begin{itemize}
\item Inconsistencies abound, and paraconsistency is the best one may aim for. In particular: 
\begin{itemize} 
\item Revealed sources are a fundamental source of the body universally quantified assertions from which reasoning must take place.
\item The totality of revealed sources is not claimed to be consistent. (Many inconsistent subsets may exist.) Informed scholars must resolve contradictions when needed.
\item Original eye-witness accounts produce evidence, and so do indirect testimonies. These may have higher priority than the 
original source facts these accounts are commenting upon. This also holds if the source facts are understood in a metaphoric fashion. Proximity in time to the causes of creation of the original sources (for instance measured by means of counting number of intermediate witnesses) increases confidence.
\item Science (including logic and mathematics) may produce valid assertions which may be inconsistent with revealed sources. Scientific fact wins out against revealed fact, moving the latter into a metaphoric status, which is then in need of explanation by interpreting scholars.
\item A improved level of knowledge of science may lead to modified assertions to which a higher degree of confidence is assigned. Ordinary resolution of inconsistencies amongst scientific results.
\end{itemize}
\item  Distributed and autonomous human judgement performed by groups of informed scholars plays and will play a major role in legal decision making which may overrule at any time  all formalized deduction from acquired database of accepted legal assertions. As a consequence:
\begin{itemize}
\item Islamic legal reasoning cannot become outdated, it is essentially a contemporary phenomenon, and,
\item the Islamic Legal Process proceeds concurrently at different locations and different courts may judge quite differently about similar cases at the same time,\footnote{This potential inconsistency of simultaneous judgements on comparable cases is a price that must be paid if the Islamic Legal Process is supposed to deliver contemporary judgements at all times.} and,
\item logic is supposed to be supportive of this mechanism and must not in any way be construed as an ``objective'' replacement of conscious group decision making by informed scholars.
\end{itemize}
\item Resolution of contradictions makes use of geographically based priorities, with local Islamic courts having more impact than distant 
ones,\footnote{The topology impacting on this geographic form of priority reasoning is in terms of ``logical'' distance between communities rather than physical (metric) distance between locations of courts.} recent 
court judgements have more relevance than older ones of the same court.\footnote{This higher priority of more recent judgements must not be confused with the fact that older assertions (assumptions) may win out against younger ones  in case these assumptions have not come about from explicit court activity. If description of fact is at stake, proximity in time is counted as an advantage.}
\item Altogether Real Islamic Logic deals with (at least) eight priority mechanisms at the same time:
\begin{itemize}
\item priority of science over revelation,
\item priority of improved science over previous scientific findings,
\item priority of confirmed interpretation (if sources are understood metaphorically) over revelation,
\item priority of direct witness reports over indirect ones,
\item priority of propositions put forward by highly regarded scholars over propositions produced by less highly regarded ones,
\item priority of propositions put forward by directly involved (concerning the issue at hand) individual scholars over judgements made by individual scholars from a more distant position,
\item priority of recent court judgements over older judgements of the same court,
\item priority of nearby (physically or community wise) court judgements over more distant ones.
\end{itemize}
As it stands human decision making is essential to balance the relative weights that must be assigned to these different priorities.\footnote{Instead
of considering logics that make use of priorities one may look at multiple criterion decision making as a related non-religious phenomenon.}

The simultaneous presence of a number of priority mechanisms renders Real Islamic Logic astonishingly complex but it constitutes no reason not to analyze its working in detail, on the contrary, it suggest that much work can be done.
\end{itemize}
\end{enumerate}

\subsection{Comparing Crescent-star Finance and Crescent-star Logic}
Crescent-star Logic stands for Real Islamic Finance stripped from its political and religious objectives. Similarly Crescent-star Finance is Islamic Finance stripped from its religious, political and ideological objectives. Both themes can be contemplated and advanced by non-muslims. There is a difference, however, because for a non-muslim Islamic Finance represents a reasonable comprehensible deviation from conventional finance and the effect of adherence to that deviation can be investigated in an impartial way both in theory by way of making use of thought experiments and in practice by means of observation of real or of artificial (that is experimental) economic processes.

Work on Crescent-star Finance may, at least in principle, reveal weak points concerning Islamic Finance that need to be taken into account by muslims pursuing Islamic Finance in its full meaning.  It may also lead to the discovery of new financial products which Islamic scholars are likely to consider morally adequate ({\em halal}).

At this early stage Crescent-star Logic is a hypothetical matter altogether because the inclusion and exclusion of formal techniques as well as philosophical methods  for Real Islamic Logic needs to be worked out from an application perspective. Nevertheless a stage can be imagined where Crescent-star Logic can be abstracted from real Islamic Logic in way comparable to the way logic programming has been obtained from programming in PROLOG.

\section{Objectives of Real Islamic Logic}
From its introduction Islam has conceived of itself as modern. Judgements produced by Islamic courts are modern in the sense that these are recent decisions taking all jurisprudence into account as well as all classical Islamic sources. Real Islamic Logic can be imagined as supportive of preservation of this form of modernity. 

The need for and role of real Islamic Logic will now be analyzed in more detail. First of all it is claimed that Islamic Finance by itself produces a significant incentive to define and investigate systems for Real Islamic Logic.

\subsection{Islamic Finance requires Real Islamic Logic}
Islamic Finance involves a very frequent use of Islamic courts for decision making about individual cases of economic behavior. But economic decision are made so frequently that the vast majority of such decisions must be taken without making use of the informed advice of an Islamic Court. Real Islamic Logic must support the reasoning taking place outside Islamic Courts. For a system of Islamic Finance it seems very useful if that form of reasoning can be understood in a scholarly fashion just like the logic of other forms of reasoning not primarily depending on human deliberation.

It is not easy to grasp why a strict adherence to the well-advertised tenets of Islamic Finance will lead to a better society. In fact all of these tenets, which can be formulated in terms of restrictions, are hard to understand in three different ways: 
\begin{itemize}
\item what the restriction means in technical terms,\footnote{Probably the clearest restriction that underlies Islamic Finance is prohibition of interests. In, \cite{BM2010} it has been pointed out that even that concept is conceptually far from obvious.} and,
\item how the restriction has come about by means of a chain of reasoning starting from original sources, and,
\item is there an expectation or intention that and observing the restriction will have positive moral or social consequences and if so, finally,
\item why compliance with the restriction might lead to a better world in moral terms.
\end{itemize}
The combined complexity and interaction of Islamic Finance renders it extremely complex with the effect that recourse to groups of Islamic scholars is regularly needed. Their judgement may be asked not only for the evaluation of generic schemes and financial products but also for the assessment of individual  plans for economic transactions. This may be exactly what is good about Islamic Finance on the long run. 

If one believes that preservation of the impact of Islamic courts is an important objective, then the development of a systematic approach to Real Islamic Logic is justified because it adds to the feasibility (improved partitioning of tasks between individual agents and courts) and to the cost reduction (reduction of court workload) of a well-organized system of Islamic Finance. That in turn may motivate research activities about Real Islamic Logic from the long term perspective of Islamic Finance.

\subsection{The role of Real Islamic Logic in more detail}
The role of Real Islamic Logic is to support methods for avoiding recourse to an Islamic court for each and every issue. Non court mediated and Real islamic Logic based decision making from distributed court output is needed to make the system sufficiently productive when scaling up. This productivity again is needed to safeguard the pivotal positions of Islamic courts which must not be overloaded with futile requests (that is requests amenable to semi-automatic resolution from the geographically distributed class of current local court outputs.)

Evidently Real Islamic Logic features a high degree of concurrency. Many agents are reasoning at the same time,  and in fact different agents will be reasoning about one-another which renders the logic epistemic. 

It is an essential part of the system that many Islamic Courts are producing their judgements (that is LP's) in a concurrent fashion. Though not democratic in a sense of elected representations the wide spread distribution of legal decision making and its commitment to reconsidering classical dilemma's must be considered a virtue.

Dealing with inconsistencies as well as dealing with novel moral dilemma's is the core of Islamic Court tasks. The Islamic legal system stands on utterly inconsistent feet and for that reason enforces a systematic dependency on human decision making. This should be considered an advantage over a system which resolves the majority of its moral dilemma's through applications of artificial intelligence and automated reasoning.

\subsection{Islamic Court Legal Proposition Processing}\label{iclpp}
Islamic Court Legal Proposition Processing (ICLPP) will be used as a descriptor of the main activity of islamic courts. The term reasoning is avoided because that  suggests regularities and methods to which ICLPP is not in any way committed. ICLPP and Real Islamic Logic are complementary mechanisms where ICLPP is by nature an entirely human group activity. It is not suggested that  a formal logic behind ICLPP must be sought. Perhaps ICLPP is best understood as multiple criterion decision making, a process which is amenable to empirical analysis but not amenable to theory based prediction by means of formalized logic.\footnote{Neural network modeling may prove a more effective technique for predicting ICLPP outcomes.}

Each court maintains a collection of previously issued (or imported) and not yet withdrawn LP's ready for application in new circumstances. A court maintains a history of new LP's as well as of withdrawals together with written accounts of the motivation for such steps. The current LP set is likely to be paraconsistent but not consistent (that is properly paraconsistent). When formulating judgements concerning new cases a court may put forward new LP's that may or may not contradict its current set of  publicly announced LP's. As a consequence older LP's may be withdrawn from the court's current LP set.  An LP in the current LP set of court $C$ may be considered a court $C$  belief. A court performs belief revision in a context of paraconsistent logic. Reasons for LP withdrawal or for the introduction of LP's that render the LP base (more) inconsistent so include:
\begin{itemize}
\item Changing viewpoints regarding the same topic (for instance because of novel insights in the effects of certain regulations, or due to the emergence of new scientific findings, or due to the, very rare, discovery of new source texts),
\item Entirely new dilemma's emerging from (changing) daily practice which require new LP's as well as revision (withdrawal) of old LP's,
\item Efforts to improve the consistence of the current LP base (with the intent to enhance non-court decision making on the its basis).
\item Paraconsistency appears because the full collection of valid (that is not yet withdrawn) LP's is likely not to be consistent at any time. Belief revision for paraconsistent theories cannot be based on the known framework tailored to classical logic, and this limitation implies (or at least suggests) that human decision making (by an IC) is essential. As it stands no rules for the production of LP's by courts are given.
Each IC therefore provides what we will call paraconsistent belief generation and revision, with these characteristics:
\begin{itemize}
\item This particular form of belief revision attempts to attach higher value to older LP's.
\item It makes use of an axiomatic basis emerging from old sources. These old sources produce closed as well as universally quantified LP's.
\item Courts may produce new universally quantified LP's. The collection of axioms thus formed is highly inconsistent. This inherited inconsistency, which is not amenable to a full resolution of conflicts, implies that very often a court will be needed to resolve a moral dilemma by means of human cognitive and emotional activity. All assertions held in Islam are essentially recent in the sense that only if an Islamic court explicitly confirms a judgement it must be held true.
\item Belief revision (which involves LP withdrawal)  is needed to allow unambiguous inference in relevant cases.
\item Inference in part consists of paraconsistent belief generation and revision, with deductive inference playing a minor role. More prominent 
inference methods are induction and reasoning by way of finding analogues and drawing corresponding conclusions.
 \end{itemize}
\end{itemize}
Courts can be called into action by individuals who conclude that Real Islamic Logic is inconclusive in a particular case, and courts may be called into action by local authorities who question the validity of Real Islamic Logic based reasoning as applied by individual muslims or by groups of muslims.

\subsection{Technical objectives and methods of Real Islamic Logic (RIL)}
RIL must provide methods of reasoning that support the rule-based and semi-automatic part of the non-court based Islamic legal process.
\begin{enumerate}
\item RIL supports reasoning based multiple criterion decision making.
\item RIL is based on a pragmatic view of ``informal logic''. Informal logic stands between philosophical logic and formal logic. It deals with the strength of arguments used in a non-formalized but yet rigorous practice. For instance in \cite{Gilbert2004} the use of emotional aspects in rigorous argument is analyzed. Emotional aspects probably cannot be avoided in the intended practice of RIL. A survey of informal logic is given in \cite{Johnson2006}, see also \cite{Groarke2007}. A convincing specimen of informal logic is found in \cite{Waller2001} where a classification of analogies is provided from a deductivist standpoint. 

There is no such thing as informal logic, instead there are different approaches to it.  In spite of that lack of clarity it seems unavoidable to place some form of informal logic in between practice and any application of formal logic. Opinions differ on the question to what extent an approach to informal logic by itself must be based on a particular selection of philosophies of science, language, and law.  It is likely that different brands of RIL may be based on different approaches to informal logic each serving as an embedding mechanism for more formalized approaches to logic.

\item RIL supports autonomous decision making on the basis of concurrent streams of current LP information from customizable collections of different courts.\footnote{Mark Burgess in \cite{Burgess2007} and in several other papers about his theory of promises, emphasizes autonomy as a vital ingredient for computer systems design, mainly because it is  scalable in principle. It is hard to imagine that autonomy will less important for the design of human powered organizations.}

A plurality of autonomously operating Islamic courts each produces in parallel an ever growing stream of LP's (legal propositions or). At each instant of time the overall legal position is determined by the family of theories reached that moment by each court.\footnote{Here paraconsistency, both for single court LP streams, and for their combination is the best one may expect because different courts may develop quite different views even when starting from the same  base theories derived from old sources.}One may think of the order of 5.000 to 10.000 courts each serving up to 150.000 persons, with new courts being installed on a regular basis wold wide.

The concurrent processing aspect of this distributed decision making process is vital. Its modeling requires formalization just as the modeling of its reasoning mechanisms requires the use of formalized logic. As an example of a concurrency model that may work in this case I mention thread algebra \cite{BergstraMiddelburg2007a} and the distributed and hierarchical versions of thread algebra which have been developed on the basis of the concurrency model in \cite{BergstraMiddelburg2007a}.
\item RIL supports separation of concerns:

\begin{itemize}
\item Well-developed and widely understood methods are needed to allow individual persons to infer LP's regarding moral dilemma's without the support of an Islamic court. This is necessary due to the number of persons that may put forward such dilemma's and the limited number of courts.

\item RIL must support real time decision making in cases where recourse to a court is self-defeating.\footnote{That may bring Short-circuit Logic
\cite{BergstraPonse2010b} into play which deals with propositional statements that allow their atomic components to change value during evaluation.}

\item RIL supports individuals and groups with making a judgement concerning whether or not a court should not be involved, and if so in 
which phase of the process.

\item RIL allows for methodical reasoning (rule based and almost or entirely automatic). Real Islamic Logic for that reason primarily provides an inference method for paraconsistent theories. 

\item RIL must support adequate documentation of all proceedings.
\end{itemize}

\item RIL allows for application specific or thematic versions. Different chapters of human activity may call for different dedicated instantiations of RIL (application specific RILs or ASRILs), allowing for the use of specialized terminology. One may think of an ASRIL for: finance, sustainability, food industry, policing, military, trade, art, politics and so on.
\item RIL involves:
\begin{itemize}
\item deduction in paraconsistent theories,\footnote{For a recent survey on paraconsistent logics written by Kees Middelburg in connection with our joint work on Islamic Finance I refer to \cite{Middelburg2011}.}
\item epistemic reasoning capabilities for supporting agents with reasoning about knowledge of themselves and of other agents,
\item logic of public announcements because all LP's produced by courts as well as all withdrawals are public,
\item deontic logic because most LP's are about obligation of prohibition of various actions or conditions,
\item priority logics because mutually inconsistent collections of LP's may be structured by means of priorities.
\end{itemize}

\item RIL makes use of a heterogeneous system of priorities where:
\begin{itemize}
\item LP's issued by nearby courts count stronger than LP's issued by distant ones (relative to a person seeking advice),
\item recent LP's by the same court have higher priority than older ones,
\item real time safety critical decisions are most likely to be taken by means of RIL,
\item RIL must indicate when a court must be called for advice.
\end{itemize}

\item RIL allows for:
\begin{itemize}
\item A flexible language to denote activities, plans, action histories and states of the world. Decision support concerning the question  whether or not a plan is permissible (and whether or not that can be asserted without the help of a court) is the primary task of RIL.
\item Integrated planning procedures that support permissible plan generation from stated goals.
\item Evaluation methods helpful to determine to what extent a past history of activities performed by one or more agents will be considered morally adequate when handed over for judgement to a specific court.
\item Heuristics for determining in which cases to refer the evaluation of past actions to an Islamic court, and
\item support for selecting a competent court when needed.
\end{itemize}
\end{enumerate}

\subsection{Feasibility of RIL and RIL oriented research}
The list of requirements for RIL is so extensive that designing an adequate and effective RIL from scratch in a single and brief project is  inconceivable. Nevertheless, such work need not be considered unrewarding only because of its ambitious appearance.

Empirical work can provide  a case base that allows for case studies with different reasoning methods. Theoretical work needs to show how the mentioned logical techniques can be brought together in a unified framework. Practical work in AI style may lead to prototypes of reasoning based decision support systems that make an RIL based reasoning engine available to a pioneering community.

As long as theoretical work aims at grouping together different techniques which supposedly must be brought together in a mature system of RIL, that work may be counted as work on Crescent-star Logic if one intends to avoid the complexities of the adjective Islamic and if one simultaneously wants to attribute the particular grouping of techniques to an intended application in RIL.\footnote{Needless to say that combined techniques may have applications outside RIL just as well and that such applications may motivate research without any incentive to make reference to either RIL or to Crescent-star Logic.}

An empirical criterion of the success of a design of RIL and supporting software is that Islamic courts promote its distribution and usage because they feel their own activities being strengthened by usage made of RIL by their client base.\footnote{This criterion is indirect in the sense that it is not explicitly dependent on any long term objectives of Islam as a movement.}

\section{Business ethics risk analysis for RIL development}
Who works towards RIL with the application perspective of a RIL enhanced ICLPP in mind faces a number of questions about morality or rather business ethics pertaining to that particular line of work. I intend to shed light on these matters in this section of the paper. First I will restate the orientation of the work on RIL for which its business ethics is to be scrutinized.

\subsection{Restating the orientation towards a RIL supported ICLPP}
A picture has been drawn of the mechanics of a scaled up Islamic Legal Process that makes use of Islamic Legal Reasoning performed within Islamic courts and of Real Islamic Logic applied outside courts with the main purpose to ease the burden of decision making for the courts.

In the absence of a well-understood Real Islamic Logic one may postulate the existence of a Natural Islamic Logic, which represents the ethical decision making practiced by muslims in circumstances  they consider sufficiently simple or sufficiently conclusive for them not to seek for advice of a court. The project to develop a Real Islamic Logic looks for more than the mere psychology of Natural Islamic Logic. The project requires a systematic approach to disentangle the reasoning  process currently performed by way of Natural Islamic Logic as to improve
the entire system at least from the point of view of acknowledged Islamic objectives.

The essential quality of the entire Islamic Legal Process is the permanent involvement of Islamic courts in many aspects of decision making, in a way that seems to resist automation. This quality has become more important  with the progress of computing technology. 

I have not made an attempt to put forward arguments that a well-designed system of Real Islamic Logic, when put in place at a grand scale, will indeed bring some particular family of relevant objectives closer to their realization. But is seems clear that a more systematic approach to that part of the legal process may foster the role of Islamic courts and for that reason it may be considered plausible that this will promote goals that motivate these courts. 

The assumption that fostering the role of Islamic courts will bring about progress towards intrinsic Islamic objectives underlies the motivation given for RIL and its application perspective. As an assumption, its validity cannot be confirmed by mere theoretical work and this assumption for that reason takes the form of an unproven axiom. Developing RIL by means of the design of a systematic approach to Natural Islamic Logic has been outlined and a survey of the techniques from formal logic that might come into play has been given. Of course the real work for developing a comprehensible and effective
Real Islamic Logic has yet to begin and a even proof of concept is missing at this stage of development. A sharp condemnation of the plan to
develop RIL cannot be based on known adverse consequences of its application.

\subsection{Working on the basis of randomized source documents}
At this point of our development of RIL and its rationale a remarkable and speculative hypothesis can be formulated. The hypothesis being that:
\begin{quote}
the entire Islamic Legal Process  inherits its powerful functionality first and for all from the fact that the heritage of the original sources, after years of evolution and codification of the outcomes of that evolution, still defeats an automated logical analysis to such an extent that:

i) the legal process is best put in the hands of a large number of concurrently operating autonomous groups of informed persons (the courts),
cooperating with a number of highly skilled and informed individuals (excellent Islamic scholars), and,

ii)  the members of these courts as well as the informed  individuals who play a role constitute a meritocracy, rather than that they have been selected by means of majority voting,

iii) properties i) and ii) are the distinctive characteristics of the Islamic Legal Process in comparison with competing mechanisms.
\end{quote}
If this hypothesis is actually valid it may also be maintained as  a plausible assumption  concerning an alternative world in which the original sources have been laid down in an almost random way. A weaker and for that reason more plausible  assessment is that after many years of autonomous evolution the Islamic Legal Process has become quite stable against (hypothetical) small modifications of the original sources. This stability against hypothetical perturbations of the source documents mainly derives from the distributed and autonomous nature of the Islamic legal process. If that process were firmly centralized each single line in the original source texts would have more impact on current decision making.

When the decision was made to found Islam around the year 720 on the basis of the collected original sources then available, the deciding actors rightly foresaw (perhaps unconsciously) that the combined complexity of these sources would defeat any attempted simplification and would keep readers and scholars busy for centuries to come, even in times of legal process automation.

\subsubsection{Centralization: a risk provoked by Islamic Finance?}
If one accepts that the distributed nature, with many autonomous courts working concurrently throughout the world, is a major strength of the Islamic legal process then the widespread call for unity in the world of Islamic Finance is seriously misguided. This may indicate a systemic risk for Islamic Finance as well as for Real Islamic Logic. Competition on the financial market may force Islamic Finance into structures of centralized control that are incompatible with  the distributed and autonomous nature of the Islamic legal process. Losing that competition, however, is problematic too because it limits the influence for which Islamic Finance has been originally set up. If one takes this risk of derailment into centralized structures seriously that gives way to another chapter of Tarek El Diwani's criticism that Islamic Finance imports western financial habits far too quickly and without due consideration of the risks thereof and of the longer term objectives of Islam (see the survey of Tarek El Diwany's positions of 
 \cite{ElDiwany1997} in \cite{Bergstra2011c}).

\subsubsection{Genetic programming needed?}
The rather mechanical picture of a thread vector of threads, each representing the behavior of a single court,\footnote{Implementing an instance of the ICLPP (see \ref{iclpp}).} operating together as a multithreaded system (in the terminology of \cite{BergstraMiddelburg2007a}) scheduled by way of strategic interleaving, fails to explain why the points of view of different courts do not diverge in time to an extent that similarities become rare. A remedy that may promote some convergence is as follows: at regular intervals a court imports randomly chosen LP's held by other courts. The probability of doing so is higher if the other court is more mainstream. To allow for this mechanism rather than merely having level 6 containing the mainstream communities of level 5 of the stratification, a degree of centrality must be imposed on level 5 communities (with the level 6 and mainstream communities being the ones for which the degree of centrality exceeds some threshold) and the more central a community is considered the more likely is an export of LP's maintained by one of its courts to an other court. 
This randomized export of LP's introduces a mechanism that may be compared to the working of genetic programming.

\subsection{Strengths, Weaknesses, Opportunities, and Threats (SWOT)}
Working towards RIL with its intended application in mind poses a researcher in a position where a wide range of moral issues needs to be dealt with. Even to develop a survey of these issues is so complicated that it cannot be satisfactorily done in this paper. But an attempt must be made. I will focus on the development of an RIL supported ICLPP as a process about which the moral consequences of significant participation are to be assessed. For this process a SWOT analysis will be made from three viewpoints, the results are rendered in a SWOT diagram which highlights key elements for each of the four categories of SWOT.

A SWOT analysis for a subject S will provide insight in the Strengths, Weaknesses, Opportunities and Threats relevant to S
More precisely a SWOT analysis is carried out with respect to three ingredients: 
\begin{description}
\item{\em subject:}  some entity, concept, or mechanism of which the SWOT is made, 
\item{\em context:} the context of a particular environment 
\item{\em perspective:} a viewpoint, or perspective from which the SWOT is formulated. 
\end{description}
For instance a SWOT can be made with subject: Financial Industry in London, context: European competition in the financial industry, perspective: economic development of the London area. An alternative perspective may be: the UK tax payer. An alternative context may be: protection of office buildings against various risks in the UK.

Two different context/perspective pairs for the same subject (development of a RIL based ICLPP) will be considered, and then a third SWOT analysis is made
with the legal process of a constitution based liberal democracy as a subject in the context of its competition with an RIL based ICLPP  and from the perspective of a supporter of the liberal democratic political structures. Together this leads to three SWOT diagrams.
\begin{enumerate}
\item RIL based ICLPP (entity)  in the competition with the current ICLPP which it eventually seeks to replace context), seen from inside Islam in its current form (perspective).
\item An RIL based ICLPP (entity) in the context of Islam (having embraced RIL and its applications perspective) competing with other non-Islamic ideologies (context), seen from inside a hypothetical stage of Islam having adapted to its modified process (perspective).
\item An RIL based ICLPP seen as an outpost of Islam (entity) in the context of a world with a number of constitution based liberal democracies for which RIL based ICLPP may be a competing process (context) seen from a classical democratic perspective outside Islam.
\end{enumerate}

A SWOT can be made from outside Islam and from inside for instance and the results that are obtained may differ vastly. Moral risks can only be formulated from a certain perspective. The three SWOT diagrams are meant to deal with three kinds of objections each rendered as weaknesses or threats  within an appropriate SWOT diagram.
\begin{enumerate}
\item Objections to development of RIL from a contemporary Islamic perspective, not particularly keen on legal process innovation.
\item Objections to the application perspective of RIL in terms of legal process innovation, given the need to compete with other systems outside Islam.
\item Objections to RIL development raised by those who consider ICLPP to constitute a risk for the maintenance of other (non-Islamic) legal processes for which the legal or ethical basis is considered superior, or at least of an independent value in need of preservation (seen from some perspective outside Islam).
\end{enumerate}


\subsubsection{A SWOT from a simulated internal perspective}
This SWOT analysis highlights the ways in which a participant to the development of RIL may be criticized by participants of the current ICLPP. Options for such criticism are listed as treats in this SWOT analysis. The perspective is in fact a (hypothetical) inside conventional ICLPP perspective which is formulated from outside Islam.\footnote{Because the author is its designer, this SWOT cannot be considered a ``conservative Islamic SWOT'', though approximating just that was intended.}
\begin{description}
\item{\em Strengths.} (i) potential new development for ICLPP, which may be in need of innovation.
\item{\em Weaknesses.} These take the form the RIL may not be developed (or used after development) to the satisfaction of its users and their local courts. Indeed RIL development is risky: it may be inconclusive, there may be scaling problems, results may be unconvincing or even
counterproductive.
\item{\em Opportunities.} (i) Constituting a new area for applied logic, thus attracting significant external interest, (ii) attracting the attention of a young generation of muslims.
\item{\em Threats.} (i) being considered innovation for its own sake, unable to keep focussed on true and classical legal foundations of Islam, (ii) 
logic becoming considered too distant from ordinary life\footnote{This is what happened with much of formal methods in computing.} (iii) scholars
constituting courts are unwilling to take notice of logical methods and for that reason unwilling to take new colleagues on board who intend to make use of such methods.\footnote{A corresponding handicap has proven almost insurmountable in software engineering.} (iv) opponents of RIL may successfully claim that introduction and usage RIL will lead to uniformity which then leads to centralization, a development that they want to prevent at all costs.
\end{description}

\subsubsection{A SWOT diagram from an friendly external perspective}
This SWOT analysis concerns the evaluation of RIL supported ICLPP (which serves as the motivation for RIL designs and development), from a (hypothetical) Islamic perspective. This perspective is formulated under the assumption that RIL supported ICLPP (or at least a development in that direction) has obtained ample support in mainstream Islam.\footnote{Of course many different ``inside perspectives'' can be thought of. This one has been designed by the author and for that reason does not really qualify as an inside perspective.}
\begin{description}
\item{\em Strengths.} (i) autonomy, (ii) real time decision making enabled about all daily matters on moral grounds, (iii) grounded in a long history, (iv) adaptation to local circumstances of a community, (v) law is not seen as an artificial fence which citizens may always try to approach (or even try to cross), rather law permanently guides everyone at in appropriate directions.
\item{\em Weaknesses.} (i) lacking cohesion (correspondence) between different courts, (ii) low authority by defective grounding of court decisions. (ICLPP may be too weak a process to produce grounds for decisions that are widely accepted), (iii) inability to develop adequate customizations (local adaptations) of the ICLPP, again with lacking authority as a consequence. (iii) Unclarity about its own ambitions: should this process be installed everywhere on the long run, or only in some countries with other countries following other governance principles? Should the process be made subordinate to other legal systems if those are (locally) dominant because a majority of a population has voted that way? If that is done, can it still retain its strength? 
\item{\em Opportunities.} (i) Real Islamic Finance may be developed on the basis of Islamic Finance and Real Islamic logic: less important restrictions may subsequently be abandoned and the courts may be much more involved in ethical decision making about particular financial strategies and activities, (ii) customization of ICLPP to various local circumstances may be improved, (iii) RIL may be developed and used to facilitate a more efficient and for that reason also more frequent use of courts. 
\item{\em Threats.} (i) centralization may become dominant with all advantages gone, (ii) centralization may creep into the system because of the modern ICT technologies, (iii) the system is outperformed by a competing system, in particular constitution based liberal democracies which do not tolerate significant local variations in legal decision making.
\end{description}

\subsubsection{A SWOT diagram from a cautious external perspective} The third SWOT analysis works from an external position, in particular from the position of the legal process of a constitution based liberal democracy, and it grasps the reflection on its own position assuming the emergence of an RIL supported ICLPP as depicted above, which is considered an emerging competitor.
\begin{description}
\item{\em Strengths.} (i) Centralized law making produces better rules with wider acceptance, easily defended against the outcome of local autonomous ICLPP's. (ii) Consistency of law throughout a wide area allows persons to be very mobile. 
\item{\em Weaknesses.} (i) Citizens take a defensive attitude against the law (even if it has been constructed in part in their own name) more often than not viewing it as an unwanted limitation of their possible behavior and endlessly trying to stretch its boundaries. (ii) Moral arguments have lost prominent status, though some issues can only be solved by means of such arguments. (iii) Minorities may feel poorly represented by the dominant legal system.
\item{\em Opportunities.} (i) Inside area $A$, the legal platform may allow for a reduced ICLPP (reduced because not permitted to overrule some constitutional principles), thus making area $A$ more attractive to Islamic citizens and allowing $A$ to make optimal use of their talents and skills. (iii) if ICLPP structures are more effective in fighting crime and other results of lacking social cohesion, such advantages can be made use of, (iv) the same holds for other pressing social problems such as taking care of the elderly.
\item{\em Threats.} (i) If clear rules are lacking the emergence of an ICLPP infrastructure in area $A$ may endanger the authority (and effectiveness) of the existing legal structure, without replacing it to the satisfaction of a voting majority, (ii) if the right to set up an ICLPP infrastructure, (on top of $A$'s existing legal system), may conflicts between both legal processes may arise and the power structures of $A$ may come under pressure from allies of ICLPP outside $A$. Such pressures may be conventional political once, but also involve wars and acts of terrorism. (iii) When ICLPP comes into prominence, court autonomy may imply severe restrictions on individual mobility to areas supervised by other courts. 
\end{description}

\subsection{Integrating the SWOT diagrams}
These SWOT's are hardly consistent. From an an external position (the current author's position) the external SWOT takes priority over both internal ones. A worker on RIL assumes that for risk analysis and for selection of strategy and tactics taking the second SWOT into account may take priority over taking the first o into account. 

That imposition of priorities does not imply a substantial weakening of the arguments which have been put forward in favor of RIL and its potential use. It only indicates operational boundary conditions. Together these SWOT analysis results provide some insight the pro's and con's of participating in RIL development for a non-muslim. I conclude that participation in RIL development is defensible for an outsider to Islam who is strongly supportive of a constitution based liberal democracy. 

The outsider's position brings with it some moral dilemma's which have not been uncovered by means of the above SWOT analysis exercises. This requires some further comments.

\subsubsection{Moral issues concerning the outsider's position}\label{MoralDs}
Given the non-Islamic, that is outsider, position to which I committed myself, claiming a preference for the outcomes of the third SWOT analysis over the second and of the second one over the first will not resolve all possible ethical questions concerning the work on this paper. I would hope to be able to write in such a way that the work becomes acceptable for readers from a very wide ideological and religious spectrum. Aiming at that, however, implies that potential worries of some potential readers must be dealt with which are absent if not foreign to other potential readers. Taking the worries of some seriously can  alienate others. 

The outsider's position is complicated by worries coming from two sides so to speak. Those who consider Islam a dangerous enemy of democracy, liberalism, and, of free expression of opinion, may regard the plan to develop logic in order to become more functional for Islam a dangerous activity (and its agent may be considered naive at best). For people from Islam this perspective is so unattractive that they may not be willing to read a paper which pays any attention to it. Still I have to ask for some attention because my outsider's position implies that such voices, often coming from so-called conservative politicians, cannot be overheard. Now what can I say? The moral issues cannot be definitively dealt with in isolation. Ethical questions about research and development must be considered in a larger context, taking corresponding problems in other fields into account.

I will make use of two fields of research and development as a tool for comparison: (i) applied logic in general and (ii) the development that has led to the implementations if Islamic Finance.

I notice that many applications of logic in computing have been ready made for application in the military sector. Such applications can be developed by one's enemy too, because publication is open to anyone who pays the relatively low cost of disclosure including an enemy. Thus even if one agrees that Islam is an ``enemy'' and that RIL can be applied in Islam, that state of affairs does not imply any moral problem for an RIL worker, the potential support for an enemy holds in both cases, and we know that right wing ideologists hardly ever suggest not to perform applied science and technology because that may empower the military and industrial capabilities of their enemies more than those of their allies. 

\subsubsection{Comparison with defense system oriented research considered flawed}
In spite of this comparison with applied logic, an ethical problem certainly emerges because RIL might be considered to be so specific for its Islamic background that working on RIL can only be useful for applications aiming towards a particular form of Jihad. Thus application asymmetry can
lead to a moral problem. Against this objection, which portrays
supporting work on RIL as less defensible than ``ordinary applied logic focusing on military applications''  the following can be said.
\begin{itemize}
\item In the initial stages RIL's development is not conclusive from an application point of view. Until that is the case there is no true specificity of RIL oriented work for applications in the context of ICLPP (an intended specificity alone implies no moral problem with regard to asymmetric applicability).
\item If RIL  is successfully developed its methods and techniques may have other applications besides the intended one (thus widely accessible publication is mandatory if an asymmetry based critique is to be avoided).
\item If RIL matures and is applied in the Islamic Legal process in the intended way that state of affairs may  lead by way of computational evolution to centralized, or rather uniformized, control of the legal reasoning used outside courts, because monitoring information is likely to end up in public data bases, thus allowing legal meta-studies to be made.\footnote{In medicine world-wide data bases with individual health information have opened new possibilities for evidence based treatments, sometimes at cost of local cultures of the medical profession.}
\item The situation may be compared with the development of Islamic Finance. Islamic windows\footnote{I will assume that a non-Islamic bank can perform Islamic activities through its Islamic Window (IW) provided key personnel of that  IW is Islamic and these staff members perceive their activities 
performed for IW as being compatible with their view of Islam.

Absent that condition an Islamic Window may fail to qualify as Islamic following the definition of that notion that has been put forward in section \ref{defOfIslam}. This is not inconsistent as the naming is done by a non-Islamic organization not committed to a high level of scrutiny concerning this part of its naming and branding strategy} of  banks managed  from UK and USA have played an important role in getting the set of Islamic products defined and in their subsequent commercialization. This matter seems not to have drawn much critical attention from fierce opponents of Islam. That lack of criticism is quite understandable: a few highly internationally operating western banks recognized a market opportunity and they took advantage of it successfully. These western banks have been so successful in that activity that it may be valued as a significant achievement that native banks from Islamic countries (that is, countries where a large part of the population is muslim)  finally have been able to acquire a leading role in matters of Islamic Finance. Sustainability of these leading roles has yet to be shown.

\item As an economic phenomenon the evolution towards prominence of native banks in providing the products of Islamic Finance (in Islamic countries) may be compared to the prominence which Japan has acquired in the car industry between, say, 1970 and 2010. There is no need to view the acquisition of the prominent role of native banks as a manifestation of Jihad because ordinary market competition can be used as an alternative and more attractive explanation of that development. Jihad was involved earlier when the preconditions of these developments were set.

\end{itemize}
These arguments together make me believe that an outsider should not (or rather may ask not to) be sharply criticized  (by other outsiders, perhaps including anti-Islam theorists) for doing work in such a way that he or she directly contributes to a development of RIL, including its intended application.

\section{God's existence  as an AI motivated hypothesis}
Making the scholarly topic of logic dependent on the unprovable and, from a certain perspective unscientific, assertion of the 
existence of God\footnote{I am assuming that a if a god exists  he is unique. Further by convention, if a god exists, God denotes that unique entity. 
If no god exists God is undefined (nonexistent). 
Now speaking of the existence of God makes sense because it may be true but it need not be true. Writing ``God'' can be done without already implying his existence. The situation may be compared with division by zero. In that admittedly much simpler case ``existence of 1/0'' is a property
which is usually considered wrong but nevertheless it is true in some structures, for instance in meadows \cite{BethkeRodenburg2010}. } seems to be a step backward, a denial of the definitive achievements of many centuries of rational thought. This problem of course disappears if God's existence can be shown after all. Through past centuries many candidate proofs have been invented and equally many refutations were designed leaving the perspective of finding such proofs rather bleak. The concept of Real Islamic Logic may shed some additional light on this matter, however, and some of its implications for the ``god hypothesis'' are discussed below.

As a disclaimer I must start with asserting is that I don't want to imply in any way that a belief in the existence of God not based on some form of proof is somehow problematic or, more specific for this paper, that such a belief leads to a point of departure from which development  and application of Real Islamic Logic is less plausible after all.

This section is meant for those readers who are so sceptic about the existence of God that this attitude virtually removes their curiosity for the topic of this paper.

First of all a proof of the existence of God may be compared to a proof of the existence of an ideal object in mathematics: the assumption of his existence can strengthen  human thought. The assumption of God's existence may provide intellectual and emotional empowerment.\footnote{Needless to say assuming the existence of a god  may also lead to the opposite effect. Without an appropriate context the presence or even dominance of that assumption in a community has unpredictable consequences.} The comparison of God with a physical or cosmological phenomenon which must be indirectly observed by means of clever experiments and perhaps though the eyes of extremely complex cosmological mathematics may be considered less attractive.

\subsection{Making the existence of a god explicit}
I will assume that God is defined to be an eternal entity equipped with a will and able and willing to reflect upon human existence through its 
entire historical development. In some way God may also influence human existence, and adequately following his will reinforces that phenomenon, a thesis upheld even in the absence of any empirical data about it. 

The use of the assumption of the existence of God lies in the possibility to strive to conduct one's life in such a way that this complies with God's will, together with the assumption that doing so is fruitful on the long run. A substantial religion incorporates a long standing cultural heritage providing its members the tools necessary and sufficient to follow God's will as best as possible according to the communal conceptions available during his or her life.

\subsubsection{Irrationality of an existence assumption denied}
At this stage one might immediately reply: there is no ground for the hypothesis of God's existence whatsoever, and the existence assumption is irrational and counterintuitive.\footnote{A famous remark by Lapace, in reply to a question posed by Napoleon reads ``Sire, je n'avais pas besoin de cette hypot\`{e}se-l\`{a}'' (Sire, I had no need for that particular hypothesis.) I will argue that a need (or at least a reasonably convincing use) for this assumption, admittedly absent for the ambitions of Laplace and for developing most of Science, may have reappeared in recent times, with the growth of technology, which asks for other mental schemes than ``classical'' scientific explanation.} Without some ``objective'' confirmation the assumption of God's existence goes against modern scientific thought. In reply to this it might be stated that looking in the sky one may not get any clue about the hypothesis of a big bang and the corresponding  beginning of physical time either, although such phenomena are nowadays considered ingredients of plausible explanations of the universe's history. God's existence, or rather the irrationality behind that assumption, may be a singularity of logic or a singularity for rational thought comparable to the big bang being a singularity of space and time. This argument, if one attributes any value to it at all, responds to the alleged irrationality of assuming the existence of God. But whereas the big bang hypothesis constitutes (nowadays) a part of a simplified picture of the universe given the need to account for a wealth of experimental data, God's hypothetical existence cannot be rationalized in the same way due to a lack of compelling and relevant data.

At this stage the parallel with scientific hypothesis formation may just fail. The virtue of the assumption of God's existence, viewed as a hypothesis in need of confirmation, need not be found in its power to simplify the resulting picture of the world and its ongoing development, but may be found instead in its power to render this picture intrinsically more complex (and for that reason more computationally intractable) than any ordinary science based model would do. 

Some thousands of years ago religion probably made the story of life much more complex than a naive and ``natural'' 
explanation of the observed facts did at that time. This very complexity made religion almost inaccessible to non-intellectuals (none of them priests) and for that reason it helped to turn religion into a method for obtaining and keeping worldly power. Indeed if some 4000 years ago a king or queen plans to build a magnificent pyramid, it may be very helpful, if not necessary, to invent a religion assuming that his or her subordinates are not already committed to any religion. And perhaps not just any religion will do, its design may need to reflect the power structures it is supposed to enforce.

\subsubsection{Deductive, inductive, conductive, abductive, and productive arguments}
That existence of God can be deduced from simpler assumptions in a non-circular fashion has become unlikely after centuries of unsuccessful 
attempts. Informal logic (see \cite{Groarke2007}) recognizes more forms of argument, however. An inductive existence argument arises if it can be 
understood as a generalization from a number of existence assertions which, at least initially, are attributed a higher degree of confidence. This line or argument seems inapplicable in the case at hand. A conductive argument accumulates a heterogeneous (thereby disabling induction) collection of arguments (each individually non-conclusive) which together are considered sufficiently convincing. Conductive arguments have been put forward by many authors. A difficulty with conductive argumentation for God's existence is that if more arguments are needed the individual arguments seems to lose impact and the net gain is limited. Abductive arguments have been dealt with above; such arguments rest upon a
positive appraisal of the explanatory value of a hypothesis (in this case the hypothesis of God's existence) which is thereafter considered validated.

Having discharged deductive, inductive, conductive, and abductive arguments for the existence of God as being less promising, the methodology of informal logic seems to leave one empty-handed on the matter. The argument that is put forward below may be classified as productive: it produces a context which is advantageous in terms of its social productivity. As such its productivity may be compared, in principle (and for the sake of this argument), to the existence axioms in axiomatic set theory for which the main argument in their favor is the extremely rich and magically attractive world of mathematics that unfolds upon their assumption.
\subsection{Virtues of the existence assumption}
Assuming the existence and uniqueness of  a god, together justifying the identifier ``God'', combined  with adherence to a substantial religion, such 
as for instance Islam, an individual gets committed to the task of permanently finding out about God's will in a changing world on the basis of a long and complex oral and written tradition. The  Islamic Legal Process is a geographically distributed mechanism for carrying out this task by way of a tradition based group process. 
\\~\\
HYPOTHESIS 1. The virtue of the Islamic Legal Process (now essentially conceived as a highly distributed process without centralized control) may be understood as follows: it is more resistant against being automated than any other human process that one may imagine.\footnote{Above 8 levels of priorities of reasoning have been distinguished as being of relevance to RIL. A formal and predictable analysis of the combined effect of these 8 priority mechanisms is unlikely soon to emerge.}  In other words it poses a maximal challenge for text oriented artificial intelligence and making in addition the assumption of God's existence ensures that something fundamentally non-mechanical is at stake (or better: is 
assumed to be at stake), thereby rendering also the intention of complete (and threatening) mechanization futile.\footnote{Art and Science also represent areas where automatic robots (or softbots) won't easily replace men.}
\\~\\
HYPOTHESIS 2. The virtue mentioned in hypothesis 1 is so significant that assuming god's existence is justified for that reason.\footnote{Remarkably hypothesis 2 is in full contradiction with the idea of apocalyptic AI as surveyed in \cite{Geraci2008}. I must admit that apocalyptic AI strikes me as quite unattractive in spite of its possible futurological merits.}
\\~\\
Although the progress of artificial intelligence has not met original expectations it is a steady process and its endpoint has by no means been reached. In forthcoming years many of today's human tasks will be turned into computer based activities. 1300 years after its initiation Islam may represent mankind's best opportunity for the definition of a competence (following God's will and finding out what that means) which will defeat attempts of automation on a sustained basis. Automation of cognitive human activity which makes human interventions redundant is an undeniable threat if only a distant one, besides being a great opportunity just as well. It is against this emerging threat that a substantial religion may choose to maintain the hypothesis of the existence of God as an intellectual weapon which may be hard to replace.

Hypothesis 2 is meaningless if hypothesis 1 is unwarranted. Progress in artificial intelligence is so fast that the possibility that a single Islamic Court is automated in such a way that an external observer has difficulties in detecting that fact and in seeing the difference with the majority of ``natural''  Islamic Courts (that is it passes a kind of customized Turing test) must be taken into account even in a near future.  The highly  distributed and autonomous form of the Islamic Legal Process, until now successfully resisting an evolution towards centralized control, together with its long term stability contributes to the credibility of hypothesis 2.

Hypothesis 1 speaks of a maximal challenge for AI and that cannot be formally true: automation of the entire human cultural development for an extended period of time poses a greater challenge. So implicit in this notion of a challenge is the requirement that a functionality of a limited scope is envisaged.

\subsection{Weaknesses of this ``proof''}
Arguing for the existence of a god along the lines of hypothesis 1 and hypothesis 2 has an interesting weakness: it is implausible that this god is unique because it is unique only given the start made by its initial source documents, in combination with the sustained effort to take all source documents into account and to avoid definitive splits in the community. The fundamental difficulty of keeping the collection of communities in $I_5$ in some way coherent reflects the difficulty of the underlying task: to follow God's will. If one were to accept an existence proof of a god along these lines of thought this existence is only unique relative to the bundle of original sources, and the prophet gains prominence with respect to the god whose message he has been transmitting. 

So why might it be useful to turn things upside down and to consider god a reality and his prophet a messenger about that reality? This counterintuitive mechanism  enforces that the Islamic community is kept coherent. As a simplifying assumption it helps to understand why sustained coherence is a plausible objective. But ``united we stand'' may now be understood as a tool for preventing on a sustained basis that intellectual and emotional power is transferred into the chips of computers. However, within less than a century this idea may be proven wrong, given the progression of artificial intelligence technology. That turn of events opens the door to a ``victory'' of apocalyptic AI (see \cite{Geraci2008}).

\section{Concluding remarks}
A project for developing Real Islamic Logic (RIL) has been outlined. A risk analysis has been sketched concerning an outsider's participation to the initial phases of the development of RIL, and these risks have been assessed as being acceptable for the author at least.

As a speculation the intractability of RIL as a reasoning system has been portrayed as an advantage which should be preserved and might be exploited. That very advantage allows a further speculative thought concerning the relevance of the assumption of God's existence for strengthening the intended applications  of RIL.

The original sources of Islam, together with a long tradition of derived works define a control code (see \cite{BergstraMiddelburg2009b,Janlert2008}) for a reasoning engine which is intractable (in computing terms). Probably, with the help of modern computer science an even more intractable control code (that is a control code leading to a less tractable reasoning system) can be designed. But for the time being there is no need for that because, in spite of its age, Islam's complexity is sufficiently high to prevent automation of its reasoning process in the near future.

\end{document}